\documentstyle[prbbib,aps]{revtex}

\begin{document}
\draft
\preprint{}
\title{
"Phase" diagram of
Izing and Heisenberg cubic clusters
}
\author{ Yu.~A.~Koksharov
}
\address{Faculty of Physics,
M.V.Lomonosov Moscow State University, 119899, Moscow, Russia }

\date{\today}
\maketitle
\begin{abstract}

We have studied the ground state of
a simple cubic magnetic cluster, which contains
a spin $s$ at each corner site.
The ground state of such cluster
depends on the competition between
nearest, next-nearest and next-next-nearest-neighbor
exchange interactions.
We have calculated "phase" diagrams
for the Izing clusters with
$\displaystyle{
s={1\over 2};1;{3\over 2};{5\over 2};\infty}$
and for the Heisenberg
cluster with
$\displaystyle{s={1\over 2};1}$.
We have found that the "phase" diagram is remarkably independent on the
$s$ value.
It is important also that the Izing "phase" diagram can be used
as a rough approximation
for the Heisenberg model.

\end{abstract}

\narrowtext

In recent years much attention has been given to magnetic
properties of molecular magnets - clusters, each containing
relatively small number of paramagnetic ions. The best known
examples of such clusters are Mn$_{12}$ and Fe$_8$, which have been
intensively studied due to quantum tunneling magnetization
phenomena.\cite{Thomas} The number of transition metal ions in molecular
paramagnetic clusters extends from 2 (in dimers) up to 30, as in
polyoxometalate Mo$_{72}$Fe$_{30}$.\cite{Muller}
A spin quantum number $s$ varies
from 1/2, e.g.  for polyolate-bridged copper clusters\cite{Pilawa},
up to 5/2 for iron and manganese clusters.

It appears that the
majority molecular magnets can be described by a model of
single-particle localized magnetic moments ("spins") coupled by
the Heisenberg exchange interactions.\cite{Bencini} In spite of smallness
of molecular magnets, a numerical calculation of the energy
spectrum could be a nontrivial task, since the dimension of the
Hilbert space for a system of $N$ spins $s$, given by $(2s+1)^N$, grows
rapidly with $N$ and $s$.
For example, in the case of Mn$_{12}$ the exact
calculation of energies of all spin states arising from the
coupling of eight $s_a=2$ spins and four $s_b=3/2$, whose number is
$(2s_a+1)^8(2s_b+1)^8 = 10^8$, is practically out side the capability
of the most powerful computers.\cite{Caneschi}

In magnetic clusters each
of spin has a few neighbors, resulting in a frustration of
exchange interactions. Because of complexity of exchange
pathways, both a sign and a magnitude of exchange constants are
difficult to predict {\it a priori}. A cluster becomes magnetic, if it
has a non-zero-spin ground state.
It is interesting to study effects of
competing interactions on the ground state of a relatively small
and highly symmetrical spin cluster, for which a wide range of
exchange parameters can be relatively easily examined.
In this paper we continue our examination of zero-temperature
magnetic properties
of cubic clusters\cite{Koksharov} and present
results of an investigation of ground-state properties of a
cubic cluster of eight spins $s$, coupled by Izing or
Heisenberg interactions.
There are enough examples of real molecular clusters,
containing metal ions at corner sites of a simple cube.
Unfortunetly, these ions often have "non-magnetic" valence
(e.g. copper-dithiolato species and related compounds\cite{Garland},
cobalt cyanide clusters\cite{Berseth}).
However, many potentially interesting cubic clusters
(for instance, lanthanide sulfido clusters\cite{Melman},
met-cars\cite{Rohmer}) remain unexplored by magnetic techniques.

A sketch of the cluster is shown in
Fig.~1.
The model Heisenberg Hamiltonian of the system takes the form
$$
H = J_1H_{nn} + J_2H_{nnn} + J_3H_{nnnn}
\eqno (1)
$$
with
$$
H_{nn}=
\vec{S}_0\vec{S}_1 +
\vec{S}_0\vec{S}_3 +
\vec{S}_0\vec{S}_4 +
\vec{S}_1\vec{S}_2 +
\vec{S}_1\vec{S}_5 +
\vec{S}_2\vec{S}_3 +
$$
$$
+\vec{S}_2\vec{S}_6 +
\vec{S}_3\vec{S}_7 +
\vec{S}_4\vec{S}_5 +
\vec{S}_4\vec{S}_7 +
\vec{S}_5\vec{S}_6 +
\vec{S}_6\vec{S}_7;
\eqno(1a)
$$
$$
H_{nnn} =
\vec{S}_0\vec{S}_2 +
\vec{S}_0\vec{S}_5 +
\vec{S}_0\vec{S}_7 +
\vec{S}_1\vec{S}_3 +
\vec{S}_1\vec{S}_4 +
\vec{S}_1\vec{S}_6 +
$$
$$
+\vec{S}_2\vec{S}_5 +
\vec{S}_2\vec{S}_7 +
\vec{S}_3\vec{S}_4 +
\vec{S}_3\vec{S}_6 +
\vec{S}_4\vec{S}_6 +
\vec{S}_5\vec{S}_7;
\eqno(1b)
$$
$$
H_{nnnn} =
\vec{S}_0\vec{S}_6 +
\vec{S}_1\vec{S}_7 +
\vec{S}_2\vec{S}_4 +
\vec{S}_3\vec{S}_5;
\eqno(1c)
$$
for the Heisenberg Hamiltonian,
and
$$
H_{nn}=
{S}^{z}_0{S}^{z}_1 +
{S}^{z}_0{S}^{z}_3 +
{S}^{z}_0{S}^{z}_4 +
{S}^{z}_1{S}^{z}_2 +
{S}^{z}_1{S}^{z}_5 +
{S}^{z}_2{S}^{z}_3 +
$$
$$
+{S}^{z}_2{S}^{z}_6 +
{S}^{z}_3{S}^{z}_7 +
{S}^{z}_4{S}^{z}_5 +
{S}^{z}_4{S}^{z}_7 +
{S}^{z}_5{S}^{z}_6 +
{S}^{z}_6{S}^{z}_7;
\eqno(1d)
$$
$$
H_{nnn} =
{S}^{z}_0{S}^{z}_2 +
{S}^{z}_0{S}^{z}_5 +
{S}^{z}_0{S}^{z}_7 +
{S}^{z}_1{S}^{z}_3 +
{S}^{z}_1{S}^{z}_4 +
{S}^{z}_1{S}^{z}_6 +
$$
$$
+
{S}^{z}_2{S}^{z}_5 +
{S}^{z}_2{S}^{z}_7 +
{S}^{z}_3{S}^{z}_4 +
{S}^{z}_3{S}^{z}_6 +
{S}^{z}_4{S}^{z}_6 +
{S}^{z}_5{S}^{z}_7;
\eqno(1e)
$$
$$
H_{nnnn} =
{S}^{z}_0{S}^{z}_6 +
{S}^{z}_1{S}^{z}_7 +
{S}^{z}_2{S}^{z}_4 +
{S}^{z}_3{S}^{z}_5;
\eqno(1f)
$$
for the Izing Hamiltonian.
Here $\vec{S}_i$ is the spin operator located at lattice site i,
and $S^{z}_{i}$ denotes its z component.
Terms $J_1H_{nn}$, $J_2H_{nnn}$ and $J_3H_{nnnn}$ describes the
NN, NNN and NNNN interactions, correspondingly.  It is important
that the Heisenberg Hamiltonian (1) commutes with the operator
of the total
spin $\vec{S}^2$ as well as its z component $S_z$, where
$
\vec{S} =
\vec{S}_0+
\vec{S}_1+
\vec{S}_2+
\vec{S}_3+
\vec{S}_4+
\vec{S}_5+
\vec{S}_6+
\vec{S}_7.
$
Therefore, the energy level
scheme consists of a number multiplets each of which can be
described by quantum numbers $S$ and $S_z$.
The number of states is
easily derived by noting that the spin states of $S_i$ form a basis
for the irreproducible representation $D_S$.
For example, in case of $s=1/2$
the 256 states must form a basis for the representation
$
D_{1/2}\times
D_{1/2}\times
D_{1/2}\times
D_{1/2}\times
D_{1/2}\times
D_{1/2}\times
D_{1/2}\times D_{1/2}=
(D_0 +D_1)\times
(D_0 +D_1)\times
(D_0 +D_1)\times(D_0 +D_1) =
14D_0 + 28D_1 + 20D_2 + 7D_3 + D_4.
$
Thus we expect 14 singlet $S=0$ states, 28 triplets with $S=1$ and so on.
The eigenfunctions of (1) can be expressed as linear combinations
of basis functions
$\varphi_m =|\ldots\uparrow_i\ldots\downarrow_k\ldots\rangle$,
where $m = \sum\limits_{i=0}^{7}a_i2^i$,
$a_i = 1$ or 0 according as the i-th spin is up ($\uparrow_i$)
or down ($\downarrow_i$).
All the eigenvalues $\epsilon_n$ and the corresponding
eigenfunctions $\phi_n$ of the Hamiltonian (1) can be found by the
exact diagonalization of a 256$\times$256 matrix
(for example, using the standard Householder method\cite{Press}).
It should be noted that the functions $\varphi_m$ are
exactly equal to the eigenfunctions of the
Izing Hamiltonian (Eqs.(1d-1f)), which
commutes with the $S_z$ operator.
Hence, the eigenvalues for the Izing model can be
easily calculated without the energy matrix diagonalization.

Let us consider the Izing cubic cluster in the case
of $s=1/2$. For each eigenfunction
$\varphi_{m}$ we have found the average values
$\langle H_{nn}  \rangle$, $\langle H_{nnn} \rangle$ and $\langle H_{nnnn}\rangle$
(see Table 1).
\vbox{
$$
\vbox{
{\strut TABLE~1. Matrix elements for the Izing model.}
\vskip 2 true pt
\halign{
\vrule \strut\hfill\  # \hfill  &
\vrule \strut\hfill\  # \hfill  &
\vrule \strut\hfill\  # \hfill  &
\vrule \strut\hfill\  # \hfill  &
\vrule \strut\hfill\  # \hfill  \vrule\cr
\noalign{\hrule}
$\langle H_{nn}  \rangle$ &
$\langle H_{nnn} \rangle$ &
$\langle H_{nnnn}\rangle$ &
$\langle S_{z}   \rangle$ &
$N$ \cr
\noalign{\hrule}
1.0    & $-$1.0 & $-$1.0 & 0 & 6 \cr
$-$1.0 & 0.0    & 0.0    & 0 & 24 \cr
0.0    & $-$1.0 & 0.0    & 0 & 24 \cr
0.0    & 0.0    & $-$1.0 & 0 & 8  \cr
$-$1.0 & $-$1.0 & 1.0    & 0 & 6 \cr
$-$3.0 & 3.0    & $-$1.0 & 0 & 2 \cr
0.5    & $-$0.5 & $-$0.5 & 1 & 24  \cr
$-$0.5 & $-$0.5 & 0.5    & 1 & 24 \cr
$-$1.5 & 1.5    & $-$0.5 & 1 & 8 \cr
1.0 & 0.0 & 0.0 & 2 & 12  \cr
0.0 & 1.0 & 0.0 & 2 & 12 \cr
0.0 & 0.0 & 1.0 & 2 & 4 \cr
1.5 & 1.5 & 0.5 & 3 & 8 \cr
3.0 & 3.0 & 1.0 & 4 & 1 \cr
\noalign{\hrule}
}}
$$
}
Here
$\langle H_{nn}\rangle=\langle \varphi_{m}|H_{nn}|\varphi_{m} \rangle$,
$\langle H_{nnn}\rangle=\langle \varphi_{m}|H_{nnn}|\varphi_{m} \rangle$,
$\langle H_{nnnn}\rangle=\langle \varphi_{m}|H_{nnnn}|\varphi_{m} \rangle$,
$N$ is the number of the wavefunctions $\varphi_{m}$
with the equal values of
$\langle H_{nn}  \rangle$,
$\langle H_{nnn} \rangle$ and
$\langle H_{nnnn}\rangle$.
Fortunately, the number of different sets of
$\langle H_{nn}  \rangle$,
$\langle H_{nnn} \rangle$,
$\langle H_{nnnn}\rangle$
is much less than the total number of the
wavefunctions. For example,
we have seventy functions $\varphi_{m}$ with $S_{z}=0$ and only
six different sets of corresponding
$\langle H_{nn}  \rangle$,
$\langle H_{nnn} \rangle$,
$\langle H_{nnnn}\rangle$ (Table~1).

Using data from Table~1 we can write analitical expression for all
Izing eigenvalues $\varepsilon_{m}$:
$$
\varepsilon_{m}=J_{1}\langle H_{nn}\rangle+
J_{2}\langle H_{nnn}\rangle+J_{3}\langle H_{nnnn}\rangle.
\eqno(2)
$$
Complete set of different $\varepsilon_{m}$ can be written as follows:
$$
\varepsilon_{1}^{(0)}=J_{1}-J_{2}-J_{3};
$$
$$
\varepsilon_{2}^{(0)}=-J_{1};\
\varepsilon_{3}^{(0)}=-J_{2};\
\varepsilon_{4}^{(0)}=-J_{3};
\eqno(3)
$$
$$
\varepsilon_{5}^{(0)}=-J_{1}-J_{2}+J_{3};\
\varepsilon_{6}^{(0)}=-3J_{1}+3J_{2}-J_{3}.
$$
$$
\varepsilon_{1}^{(1)}=0.5(J_{1}-J_{2}-J_{3});\
$$
$$
\varepsilon_{2}^{(1)}=0.5(-J_{1}-J_{2}+J_{3});
\eqno(4)
$$
$$
\varepsilon_{3}^{(1)}=0.5(-3J_{1}+3J_{2}-J_{3}).
$$
$$
\varepsilon_{1}^{(2)}=J_{1};\
\varepsilon_{2}^{(2)}=J_{2};\
\varepsilon_{3}^{(2)}=J_{3}.
\eqno(5)
$$
$$
\varepsilon_{1}^{(3)}=0.5(3J_{1}+3J_{2}+J_{3}).
\eqno(6)
$$
$$
\varepsilon_{1}^{(4)}=3J_{1}+3J_{2}+J_{3}.
\eqno(7)
$$
In Eqs.(2-6) the upper index shows $S_{z}$ absolute value, the lower
index numerates energy levels with the same $S_{z}$.

The comparative analysis of Eqs.(3,4) as well as
Eqs.(6,7) immediately
shows that  eigenstates with $S_{z}=\pm 1\ {\rm or}\ S_{z}=\pm 3$
can not be the ground states
It results from Eqs.(3,5,7) that the eigenstates with $S_{z}=\pm 2$ can not
be the ground states either.

Indeed, $\varepsilon_{1}^{(2)}>\varepsilon_{2}^{(0)}$, if $J_{1}>0$.
Futher, if $J_{1}<0$,
there is a competition between
$\varepsilon_{1}^{(2)}$, $\varepsilon_{i}^{(0)}$ ($i=\overline{1,6}$) and
$\varepsilon_{1}^{(4)}$.
Assuming $J_{1}=-1$ and writing
the system of inequalities
$$
\cases{
\varepsilon_{i}^{(0)}\geq -1\ \ (i=\overline{1,6}); & \cr
\varepsilon_{1}^{(4)}\geq -1; &\cr}
$$
we get
$$
\cases{
J_{2}\in \emptyset; & \cr
J_{3}\in \emptyset. &\cr}
$$
Hence, the energy levels with $\varepsilon_{1}^{(2)}$
can not be lowest.

Analogously, we have from Eqs.(3,5,7) the relations for
$\varepsilon_{2}^{(2)}$  and $\varepsilon_{3}^{(2)}$:

\noindent
$\varepsilon_{2}^{(2)}>\varepsilon_{3}^{(0)}$, if $J_{2}>0$,
and $\varepsilon_{2}^{(2)}>\varepsilon_{6}^{(0)}\
{\rm or}\ \varepsilon_{2}^{(2)}>\varepsilon_{1}^{(4)}$, if $J_{2}<0$;

\noindent
$\varepsilon_{3}^{(2)}>\varepsilon_{4}^{(0)}$, if $J_{3}>0$; and
$\varepsilon_{3}^{(2)}>\varepsilon_{5}^{(0)}\
{\rm or}\ \varepsilon_{3}^{(2)}>\varepsilon_{1}^{(4)}$, if $J_{3}<0$.

Using
Eqs.(3,7) we can find the "phase" boundaries between
the ground states with $S_{z}=0$ and $S_{z}=\pm 4$:
$$
J_{3}=-J_{1}-2J_{2};\
J_{3}=-3J_{1}-4J_{2};\
J_{3}=-4J_{1}-3J_{2};
$$
$$
J_{3}=-1.5J_{1}-1.5J_{2};\
J_{2}=-J_{1};\
J_{3}=-3J_{1}.
\eqno(8)
$$
"Phase" diagram for the Izing cubic cluster with $s=1/2$
is shown in Fig.2.
We use the word "phase" for convenience to distinguish between regions
with different ground states.

Now we consider the Izing cubic clusters with $s> 1/2$.
The analysis for the classical spins ($s=\infty$) is the
same as for the $s=1/2$ since there are only two spin projections in
both cases.
The analitical examination of clusters with
$\displaystyle{s=1;{3\over 2},{5\over 2}}$ is simple but cumbersome since
there are much more unique sets of parameters
$\langle H_{nn}  \rangle$, $\langle H_{nnn} \rangle$,
$\langle H_{nnnn}\rangle$
(66, 129 and 999 for the eigenstates with $S_{z}=0$ and
$\displaystyle{s=1;{3\over 2},{5\over 2}}$, correspondingly).
So, we have performed numerical calculations and found that
for all values $s$
the "phase" diagram is exactly the same as shown in Fig.2.

It is interesting to compare the ground state properties of the
Izing and Heisenberg cubic clusters.
Below the case of $s=1/2$ is described.\cite{Koksharov}
For the Heisenberg cluster the zero-temperature two-spin
correlation functions have been calculated using the relation
$\omega_{ij} = \langle\phi_0|\vec{S}_i\vec{S}_j|\phi_0\rangle$,
where $\phi_0$ is the ground-state eigenfunction.
The cubic symmetry allows us to take into account only three
correlation functions:
$\omega_{01}$, $\omega_{02}$ and
$\omega_{06}$.

First we assume $J_3 = 0$ in Eq.(1).
There are four different cases to be considered:
(a) all the interactions are ferromagnetic
($J_1 < 0$; $J_2/J_1=\alpha > 0$);
(b) all the interactions are antiferromagnetic
($J_1 > 0$; $\alpha > 0$);
(c) the NN interactions are antiferromagnetic,
the NNN interactions are ferromagnetic
($J_1 > 0$; $\alpha < 0$);
(d) the NN interactions are ferromagnetic,
the NNN interactions are antiferromagnetic
($J_1 < 0$; $\alpha < 0$).

The case (a) is trivial:
the ground-state has $S = 4$,
all correlation functions are
equal to 0.25 irrespective of $\alpha$.
In the case (b) for all values of $\alpha$
the ground state is non-magnetic ($S = 0$), but both
signs and magnitudes of the correlation functions are
sensitive to the $\alpha$ variation (Fig.3).
For example, $\omega_{06}$ changes the sign
at $\alpha = 0.50$ and has the highest value 0.25
at $\alpha=1$. The latter point corresponds to
the triplet dimerized state with the eigenfunction
$\phi = (06)(17)(24)(35)$, where
$(ij) = {1\over \sqrt{2}}
(\uparrow_i\downarrow_j + \uparrow_i\downarrow_j)$.
With $\alpha\to +\infty$,
$\omega_{02}$ tends to $-0.25$,
$\omega_{01}$ and $\omega_{06}$ approach zero (Fig.3).

In the case (c) the ground state is always non-magnetic ($S = 0$),
NN and NNN interactions do not
really compete (Fig.1), since the opposite direction of NN spins
(0 and 1, 1 and 2, {\it etc}) is favorable for the
identical direction of NNN spins
(0 and 2, 1 and 3, {\it etc}).
As a result, $\omega_{02}$ is always positive
($\approx 0.25$), $\omega_{01}$
and $\omega_{06}$
are negative and depend only slightly on $\alpha$.

The case (d) is the most interesting.
Fig.4 shows the $\alpha$-dependence of the correlation functions.
If $\alpha$ is
small, all spins correlate ferromagnetically
($\omega_{01} = \omega_{02} = \omega_{06} = 0.25$).
At $\alpha_4 = -0.33$ the correlation
functions change suddenly, so that $\omega_{02}$ and
$\omega_{06}$ become negative and $\omega_{01}$ is
lowered by a factor 2.5.
It should be noted that $|\alpha_4| = 1/3$
with an accuracy of $10^{-16}$.
We found that if $|\alpha| < 1/3$ the ground state has
$S=4$. In the case of $|\alpha| > 1/3$ the ground state
becomes non-magnetic ($S=0$).

So far it has been assumed that $J_3 = 0$.
If the NNNN interactions exist,
an analysis becomes more intricate
since two degrees of freedom appear ($\alpha$ and
$\beta = J_3/J_1$).
Fig.5a shows a "phase" diagram in the case of ferromagnetic
NN interactions ($J_1 = -1$) for
$-3<\alpha<3$, $-3<\beta<3$.
The line AG separate regions with magnetic ($S=4$) and non-magnetic
($S=0$)
ground states.
The curve AG
tends asymptotically to lines
$J_2 = |J_1|$
and $J_3 = 3|J_1|$ (Fig.5a).
Hence, if either $J_2 \geq |J_1|$ or $J_3 \geq 3|J_1|$ the ground state
is always non-magnetic.
Ferromagnetic correlations between spins,
which exist in the magnetic "phase" in the case of
$J_2 < 0$ and $J_3 < 0$,
can be destroyed by means of a gradual increasing
of $J_2$ and/or $J_3$.
Let us consider, for
example, the case $J_1 = J_2 <0$, $ \beta< 0$.
If $|\beta| < 1.5$, the ground state is magnetic ($S=4$) and
$\omega_{01} = \omega_{02} = \omega_{06} = 0.25$.
When $|\beta|$ exceeds 1.5 the ground state becomes non-magnetic (S=0).
We found that the straight line $BC$ (Fig.4a),
which satisfies $J_1 = J_2 = -(2/3)J_3 <0$, corresponds to a
dimer ground state with the eigenfunction
$\phi_{0,d} = [06][17][24][35]$,
where
$[km] = {1\over \sqrt{2}}
(\uparrow_k\downarrow_m - \uparrow_m\downarrow_k)$.
For the dimer ground state the correlation
between the NN and NNN spins reduces to zero
($\omega_{01} = \omega_{02} = 0$), the NNNN spins are coupled
antiferromagnetically ($\omega_{06} = -0.75$).
An another
specific non-magnetic ground state, corresponding to the line DE (Fig.4a),
appears when $J_1 = J_3 <0$.
If $\alpha < -0.66$ the tetramerization takes place.
Namely, the non-zero correlation exists between only the
NNN spins: $\omega_{01} = \omega_{06} = 0$; $\omega_{02}\neq 0$.
The average value of $\omega_{02}$ is equal to $-0.25$ for
the four-fold degenerate tetramerized ground state
(see details below).
Obviously, the non-magnetic tetramer or dimer ground states
can be achieved also in a different way, simply in limiting cases
$J_2\to +\infty$ or $J_3\to +\infty$.

The "phase" diagram for $J_1 = +1$ is shown in Fig.5b.
The ground state has always
$S=0$ if $\alpha>-1$ or $\beta>-3$.
The line KM
tends asymptotically to lines
$J_2 = -J_1$
and $J_3 = -3J_1$ (Fig.5b)
and serves as
the boundary between magnetic (S=4,
$\omega_{01}=\omega_{02}=\omega_{06}=0.25$) and
non-magnetic (S=0) ground states.
The dimer ground state with the eigenfunction
$\phi_{0,d}$ occurs
in case of $\alpha = 1$, $\beta > 1$.
The tetramer ground state
($\omega_{01} = \omega_{06} = 0$; $\omega_{02}\neq 0$)
takes place if $\beta = 1$, $\alpha > 1$.

Now we can compare the "phase" diagrams
for the Heisenberg and Izing models.
"Phase" boundaries for
the cubic s=1/2 Izing cluster are shown in Fig.5
by dotted lines.
It is important that
the Izing "phase" diagram can be used as a rough approximation
for the Heisenberg model (Fig.5).
In particular, the asymptotic straight lines  for the
curves AG and KM (Fig.5)
coincide with the Izing "phase" boundaries.
The same situation takes plase in the case of $s=1$.\cite{Koksharov1}

We have found that the dimer
or tetramer ground state appears only if $\alpha=1$ or $\beta=1$,
correspondingly.
This regularity can be explained in the following way.
It is straightforward to show that
$H_{nn}=\vec{S}_{0257}\vec{S}_{1346}-H_{nnnn}$ and
$H_{nnn}=0.5(\vec{S}^2_{0257}+\vec{S}^2_{1346})-4s(s+1)$,
where $\vec{S}_{0257}=\vec{S}_0+\vec{S}_2+\vec{S}_5+\vec{S}_7$,
$\vec{S}_{1346}=\vec{S}_1+\vec{S}_3+\vec{S}_4+\vec{S}_6$, $s=1/2$.
Since $\vec{S}=\vec{S}_{0257}+\vec{S}_{1346}$,
we get for $\alpha=1$
$$
H=J_1[0.5S^2-3+(\beta-1)H_{nnnn}].
\eqno (9)
$$
We present Eq.(9) for $s=1/2$, in the case of $s>1/2$
the numerical parameters should be modified, but the structure of
the the equation is universal.

Assuming $J_1>0$,
it is evident from Eq.(9) that the dimerization can be
favorable only
for $\beta>1$,
because of $\langle\phi_{0,d}|H_{nnnn}|\phi_{0,d}\rangle=-3<0$
and $S=0$.
If $J_1<0$, we have to compare
$\langle \phi_{f}|H|\phi_{f}\rangle$ and
$\langle\phi_{0,d}|H|\phi_{0,d}\rangle$, where
$\phi_{f}
=\varphi_{255}$,
and $H$ is given by Eq.(9).
Taking into account that $\langle\phi_{f}|H_{nnnn}|\phi_{f}\rangle=1$,
we have the simple equation $(0.5\cdot 20-3)+1\cdot(\beta-1)=-3-3(\beta-1)$,
which results to $\beta=-1.5$.

In the case of $\beta=1$, Eq.(1) transforms to
$$
H=J_1[\vec{S}_{0257}\vec{S}_{1346}+
0.5\alpha({S}^2_{0257}+{S}^2_{1346})-3\alpha]=
$$
$$
=J_1[0.5S^2+0.5(\alpha-1)({S}^2_{0257}+{S}^2_{1346})-3\alpha].
\eqno (10)
$$
For all four eigenfunctions $\phi_{t,i}$ ($i=\overline{1,4}$)
of the tetramerized ground state
we have got
$$\langle \phi_{t,i}|{S}^2_{0257}|\phi_{t,i}\rangle=
\langle \phi_{t,i}|{S}^2_{1346}|\phi_{t,i}\rangle=0.$$
Since $\langle\phi_{f}|{S}^2_{0257}|\phi_{f}\rangle=
\langle\phi_{f}|{S}^2_{1346}|\phi_{f}\rangle=6$,
in the case of $J_1<0$ the tetramer
state has the lowest energy if
$(0.5\cdot 20)+6\cdot(\alpha-1)-3\alpha< -3\alpha$, which leads to
$\alpha<-{2\over3}$. Assuming $J_1>0$,
it is easily seen from  Eq.(10) that
the tetramer ground state is energetically preferable if $\alpha>1$.
Otherwise, states with the non-zero averages of ${S}^2_{0257}$
and ${S}^2_{1346}$ are lowest.

The Hamiltonian in (10) commutes with ${S}^2_{0257}$ and ${S}^2_{1346}$.
Hence $\phi_{t,i}$ are the eigenfunctions of these operators with
the eigenvalues ${S}_{0257}={S}_{1346}=0$.
To get more information about the tetramer ground state
we have to introduce the dimer spin operator
$\vec{S}_{02}=\vec{S}_0+\vec{S}_2$
with eigenfunctions $\phi_{02}(S_{02},M)$,
which are defined by the equations
$\vec{S}^{2}_{02}|\phi_{02}(S_{02},M)\rangle=
S_{02}(S_{02}+1)|\phi_{02}(S_{02},M)\rangle$ and
$S_{02,z}|\phi_{02}(S_{02},M)\rangle=
M|\phi_{02}(S_{02},M)\rangle$ ($S_{02}=0;1; M=0;\pm 1$).
The similar relations are valid for the $\vec{S}_{57}$,
$\vec{S}_{13}$ and $\vec{S}_{46}$ operators.
We have found that
$$
\langle \phi_{t,i}|S_{02,z}|\phi_{t,i}\rangle=
\langle \phi_{t,i}|S_{57,z}|\phi_{t,i}\rangle=
$$
$$
=\langle \phi_{t,i}|S_{13,z}|\phi_{t,i}\rangle=
\langle \phi_{t,i}|S_{46,z}|\phi_{t,i}\rangle=0; i=\overline{1,4}
$$
$$
\langle \phi_{t,1}|\vec{S}^2_{02}|\phi_{t,1}\rangle=
\langle \phi_{t,1}|\vec{S}^2_{57}|\phi_{t,1}\rangle=
$$
$$
=\langle \phi_{t,1}|\vec{S}^2_{13}|\phi_{t,1}\rangle=
\langle \phi_{t,1}|\vec{S}^2_{46}|\phi_{t,1}\rangle=0;
$$
$$
\langle \phi_{t,2}|\vec{S}^2_{02}|\phi_{t,2}\rangle=
\langle \phi_{t,2}|\vec{S}^2_{57}|\phi_{t,2}\rangle=
$$
$$
=\langle \phi_{t,2}|\vec{S}^2_{13}|\phi_{t,2}\rangle=
\langle \phi_{t,2}|\vec{S}^2_{46}|\phi_{t,2}\rangle=2;
$$
$$
\langle \phi_{t,3}|\vec{S}^2_{02}|\phi_{t,3}\rangle=
\langle \phi_{t,3}|\vec{S}^2_{57}|\phi_{t,3}\rangle=
$$
$$
=\langle \phi_{t,4}|\vec{S}^2_{13}|\phi_{t,4}\rangle=
\langle \phi_{t,4}|\vec{S}^2_{46}|\phi_{t,4}\rangle=\kappa;
$$
$$
\langle \phi_{t,4}|\vec{S}^2_{02}|\phi_{t,4}\rangle=
\langle \phi_{t,4}|\vec{S}^2_{57}|\phi_{t,4}\rangle=
$$
$$
=\langle \phi_{t,3}|\vec{S}^2_{13}|\phi_{t,3}\rangle=
\langle \phi_{t,3}|\vec{S}^2_{46}|\phi_{t,3}\rangle=2-\kappa,
$$
where the factor $\kappa$ is fractional and $\alpha$-dependent.
Explicit forms for $\phi_{t,3}$ and $\phi_{t,4}$
also depend on the $\alpha$ value.
The functions $\phi_{t,1}$ and $\phi_{t,2}$
can be always written as
$$\phi_{t,1}=\phi_{02}(0,0)\phi_{57}(0,0)
\phi_{13}(0,0)\phi_{46}(0,0);$$
$$\phi_{t,2}= {1\over 3}
\Big\{\phi_{02}(1,1)\phi_{57}(1,-1)
+\phi_{02}(1,-1)\phi_{57}(1,1)-
$$
$$
\phi_{02}(1,0)\phi_{57}(1,0)\Big\}\times
\Big\{-\phi_{13}(1,0)\phi_{46}(1,0)+
$$
$$
+\phi_{13}(1,1)\phi_{46}(1,-1)+
\phi_{46}(1,1)\phi_{13}(1,-1)\Big\}.
$$

It is easy to check  that
$
\langle \phi_{t,1}|\vec{S}_0\vec{S}_2|\phi_{t,1}\rangle=-0.75;
$
$
\langle \phi_{t,2}|\vec{S}_0\vec{S}_2|\phi_{t,2}\rangle=0.25.
$
Our calculations have showed also that always
$
\sum_{i=1}^{4}
\langle \phi_{t,i}|\vec{S}_0\vec{S}_2|\phi_{t,i}\rangle=-1,
$
and the avarage value of $\omega_{02}$ for the tetramer state is
equal to $-0.25$.

In conclusion, we have studied effects of competing exchange
interactions on the ground state of
the cubic Izing clusters with
$\displaystyle{
s={1\over 2};1;{3\over 2};{5\over 2};\infty}$.
The "phase" diagram for all Izing clusters is universal.
In case of $s=1/2$ and $s=1$ we have studied both Izing and Heisenberg
clusters.
For the Heisenberg $s=1/2$ cluster
it has been found that
the ground state can be either magnetic with $S = 4$
or non-magnetic with $S = 0$.
Intermediate values of the total spin S are not
realized for the ground state.
For $s=1/2$ the non-magnetic dimer ground state appears in two cases:
if $J_1 = J_2 < 0$; $\beta < -{3\over 2}$
and if $J_1 = J_2 > 0$;
$\beta > 1$.
The tetramer four-fold degenerate ground state
takes place
if $J_1 = J_3 < 0$; $\alpha < -{2\over 3}$ and $J_1 = J_3 > 0$;
$\alpha > 1$.
The two-spin correlation functions are much more
sensitive to $\alpha$ and $\beta$ changes
than the total spin $S$ of the ground state.
It is important that for $s=1/2$ and $s=1$ (and, probably, for 
higher spin values)
the Izing "phase" diagram can be used as a rough approximation
for the Heisenberg model.

\begin{figure}
\caption{
A sketch of a cubic cluster with eight spins.
}
\label{Fig.1}
\end{figure}

\begin{figure}
\caption{
The Izing cubic cluster.
Dotted lines show "phase" boundaries
between the ground states with $S_{z,max}=\pm 8s$
($\pm 4$ for $s=1/2$, $\pm 8$ for $s=1$, etc.)
and $S_z=0$.
}
\label{Fig.2}
\end{figure}

\begin{figure}
\caption{
The Heisenberg cluster with $s=1/2$.
The $\alpha$-dependence of the two-spin correlation functions:
$J_1 >0 $; $J_3 = 0$.
}
\label{Fig.3}
\end{figure}

\begin{figure}
\caption{
The Heisenberg cluster with $s=1/2$.
The $\alpha$-dependence of the two-spin correlation functions:
$J_1 <0 $; $J_3 = 0$.
}
\label{Fig.4}
\end{figure}

\begin{figure}
\caption{
The Heisenberg cluster with $s=1/2$.
"Phase" diagrams for (a) $J_1 <0$; $-3<\alpha<3$; $-3.5<\beta<3$, and
(b) $J_1 >0$; $-3.5<\alpha<2$; $-5.5<\beta<2$.
Dotted lines show boundaries between the ground states with $S_z=\pm 4$ and
$S_z=0$ for the Izing model.
}
\label{Fig.5}
\end{figure}


\begin{references}

\bibitem{Thomas} L.Thomas, F.Lionti, R.Ballou, D.Gatteschi,
R.Sessoli, and B.Barbara,
Nature {\bf 383}, 145 (1996),
and references therein.

\bibitem{Muller}
A.Muller, S.Sarkar, S.Q.N.Shah, H.Bogge, M.Schmidtmann,
Sh.Sarkar, P.Kogerler, B.Hauptfleisch,
A.X.Trautwein, and V.Schunemann, Angew.Chem.Int.Ed. {\bf 38} 3238 (1999).

\bibitem{Pilawa} B.Pilawa and J.Schuhmacher, J.Phys.: Condensed Mater. {\bf 8} 1539 (1996).

\bibitem{Bencini} A.Bencini and D.Gatteschi,
{\it Electron Paramagnetic Resonance of exchange coupled systems}
(Springer,
Berlin, Heidelberg), 1990.

\bibitem{Caneschi} A.Caneschi, T.Ohm, C.Paulsen, D.Rovai, C.Sangregorio and
R.Sessoli, J.Magn.Magn.Mater. {\bf 177-181} 1330 (1998).

\bibitem{Koksharov} Yu.A.Koksharov and A.N.Safronov
"Effect of competing exchange interactions on ground state of cubic $s=1/2$
Heisenberg cluster" (submited to PRB 09.04.2001,BDJ801)


\bibitem{Garland} M.T.Garland, J.-F.Halet, and J.-Y.Saillard, Inorg.Chem.
{\bf 40} 3342 (2001),
and references therein.

\bibitem{Berseth} P.A.Berseth, J.J.Sokol, M.P.Shores, J.L.Heinrich, and
J.R.Long, J.Am.Chem.Soc. {\bf 122} 9655 (2000),
and references therein.

\bibitem{Melman} J.A.Melman, T.J.Emge, and
G.G.Brennan, Chem.Commun. 2269 (2000).

\bibitem{Rohmer} M.-M.Rohmer, M.B\'{e}nard, and
G.-M.Poblet, Chem.Rev. {\bf 100} 495 (2000).

\bibitem{Press}
W.H.Press, S.A.Teukolsky, W.T.Vetterling, and B.P.Flannery,
{\it Numerical Recipes in C}
(Cambridge University Press), 1995.

\bibitem{Koksharov1} Yu.A.Koksharov, to be published

\end{references}
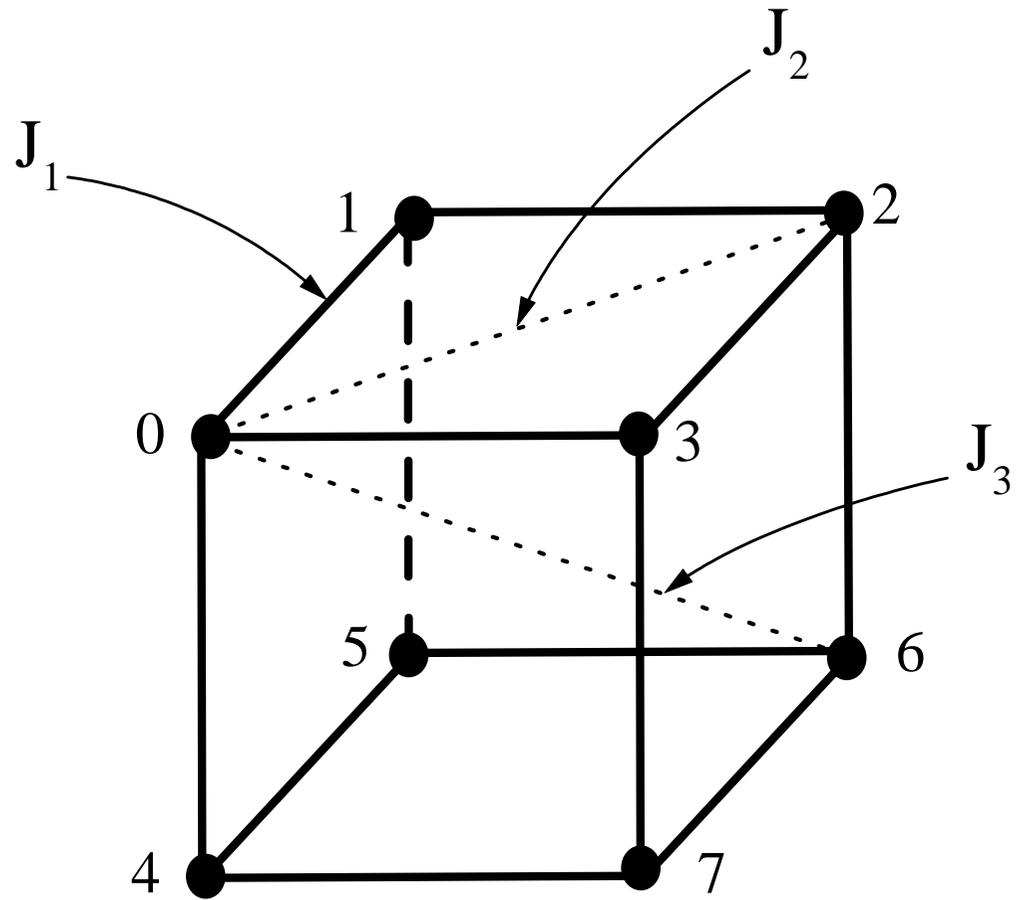
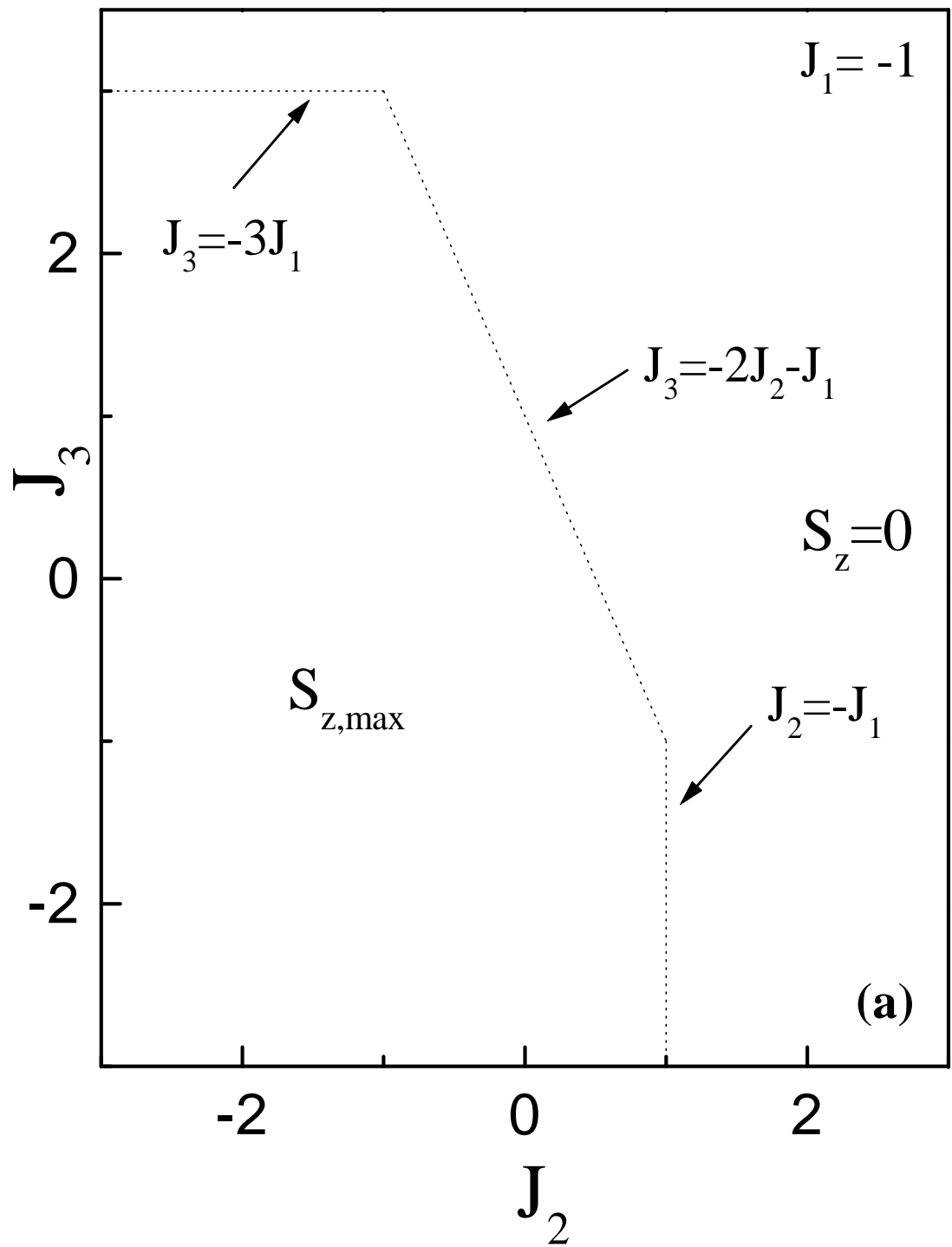
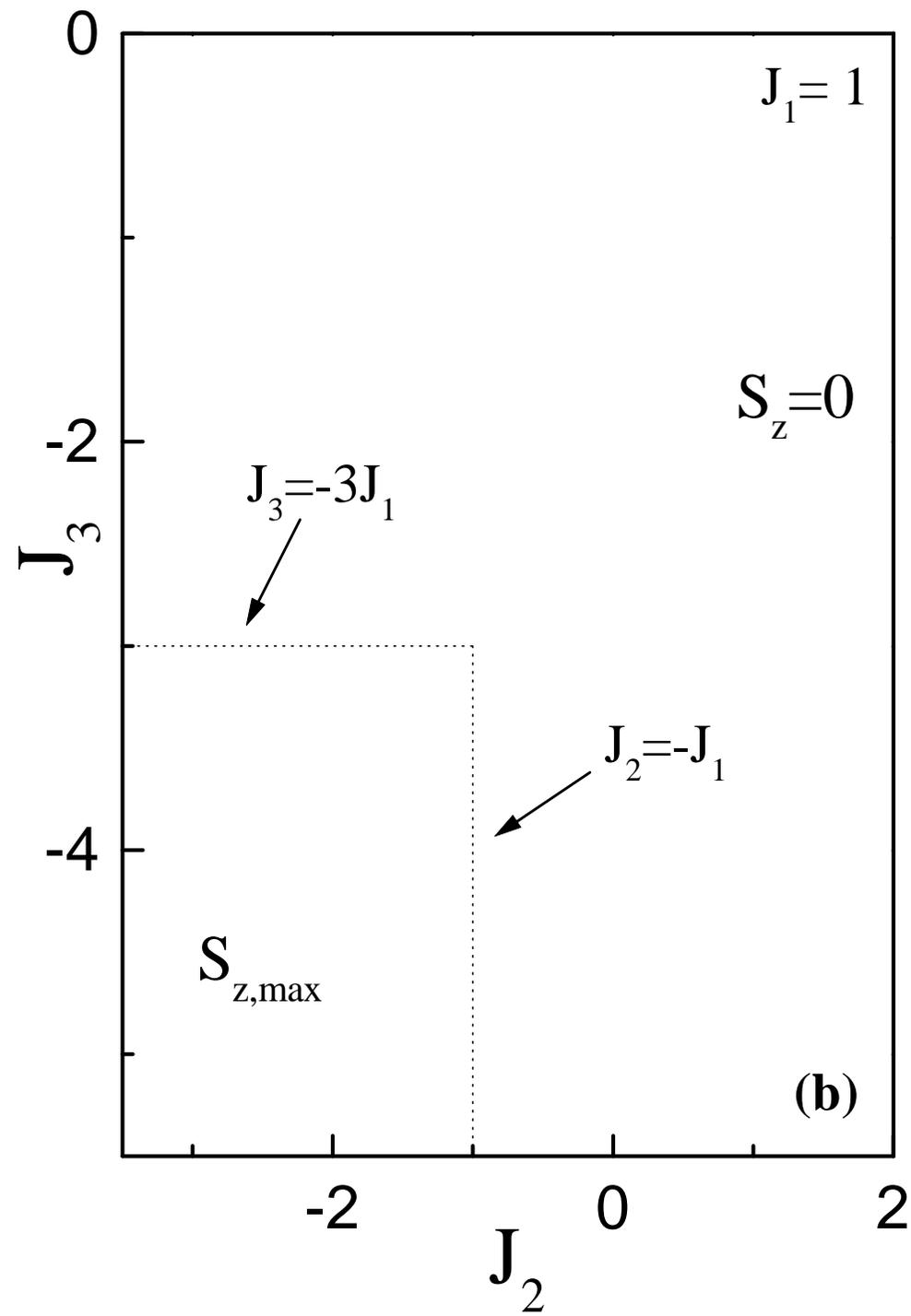
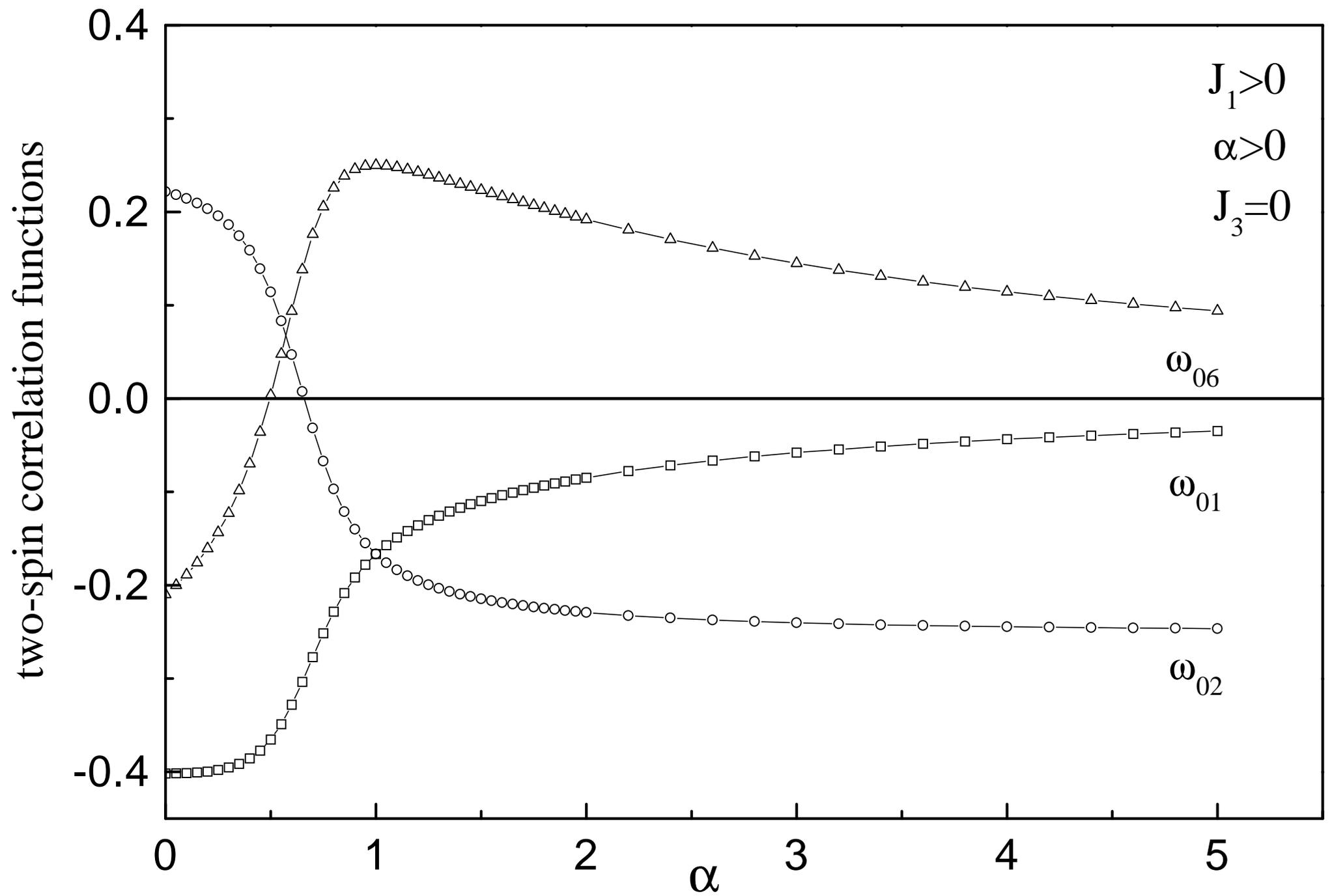
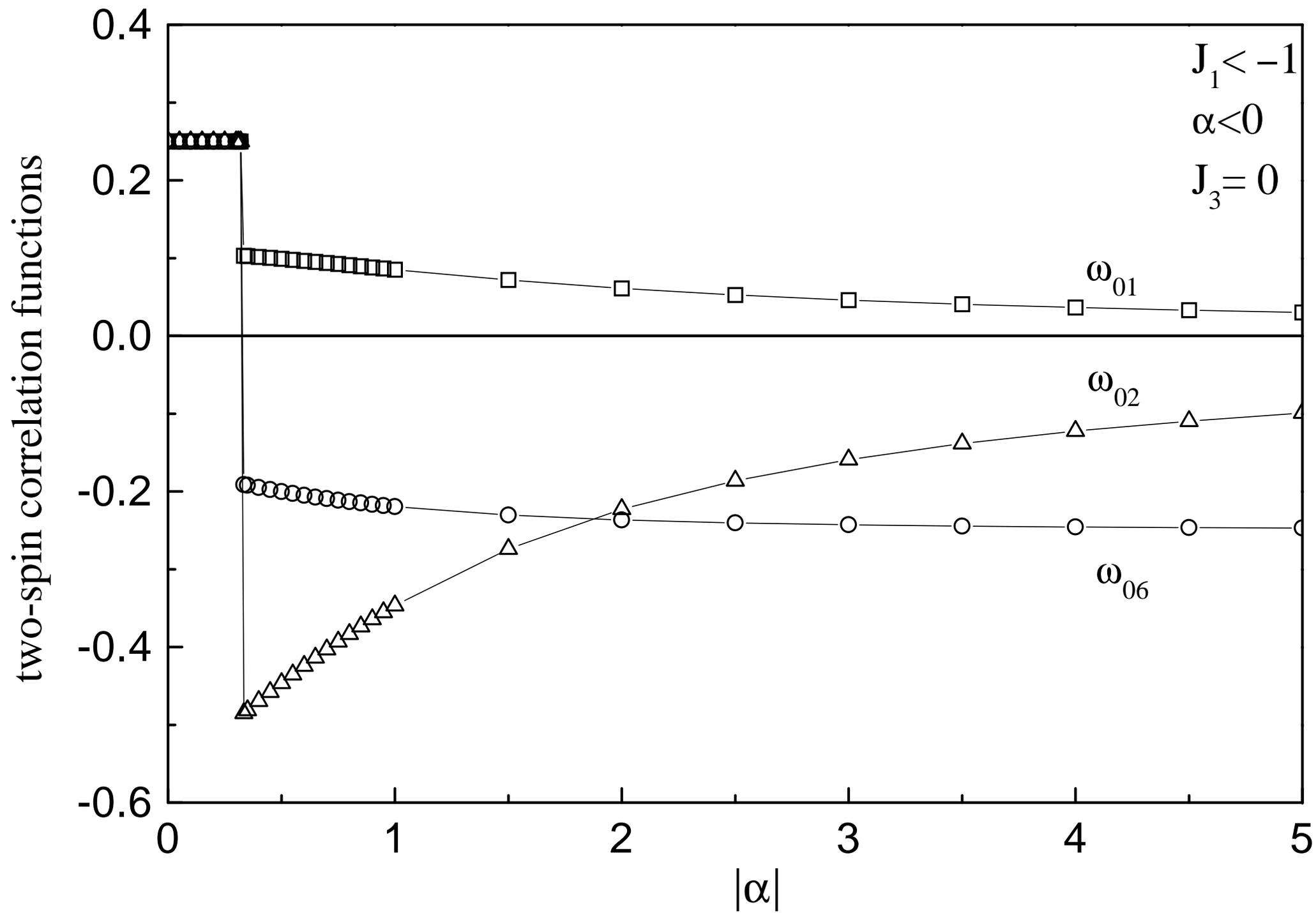
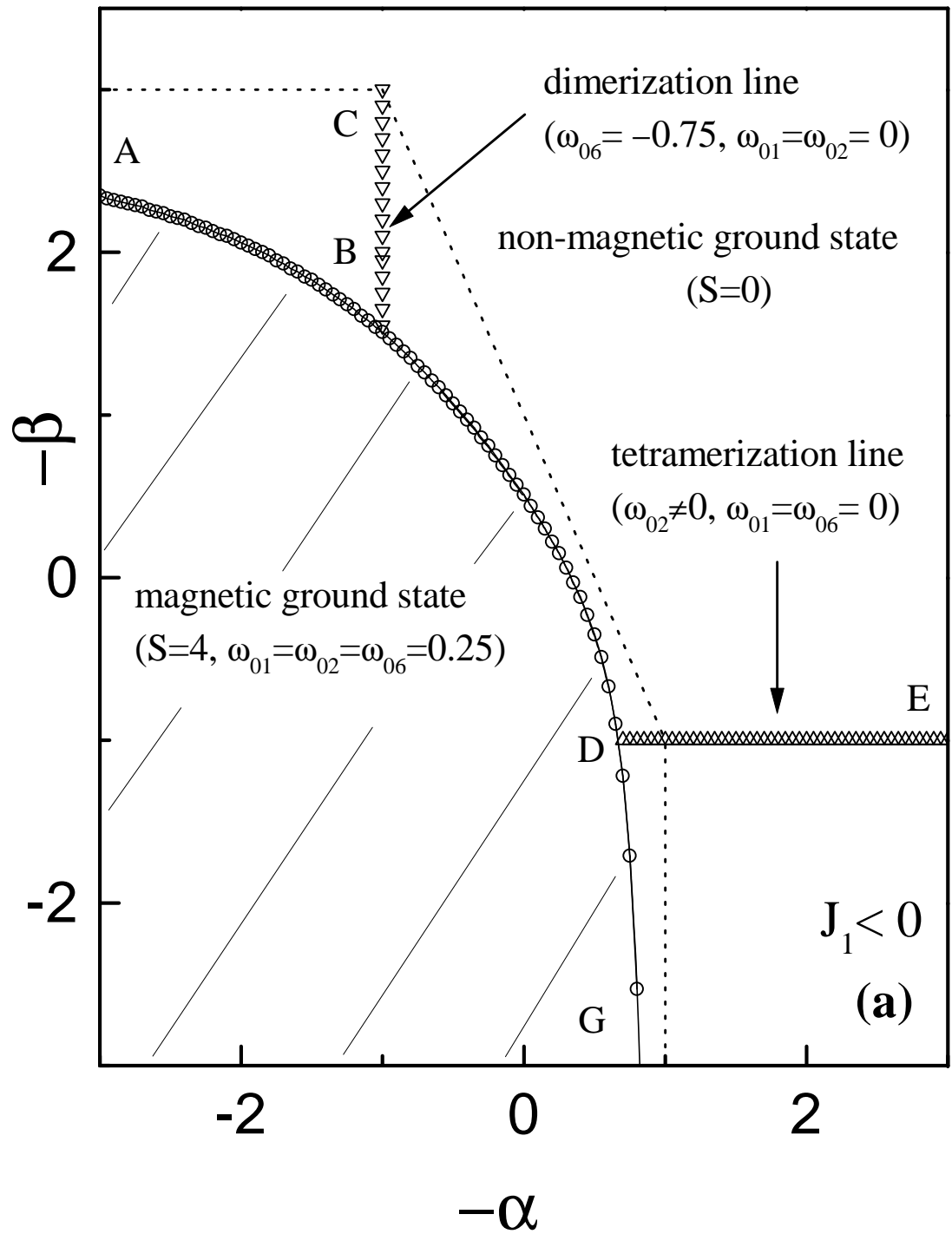
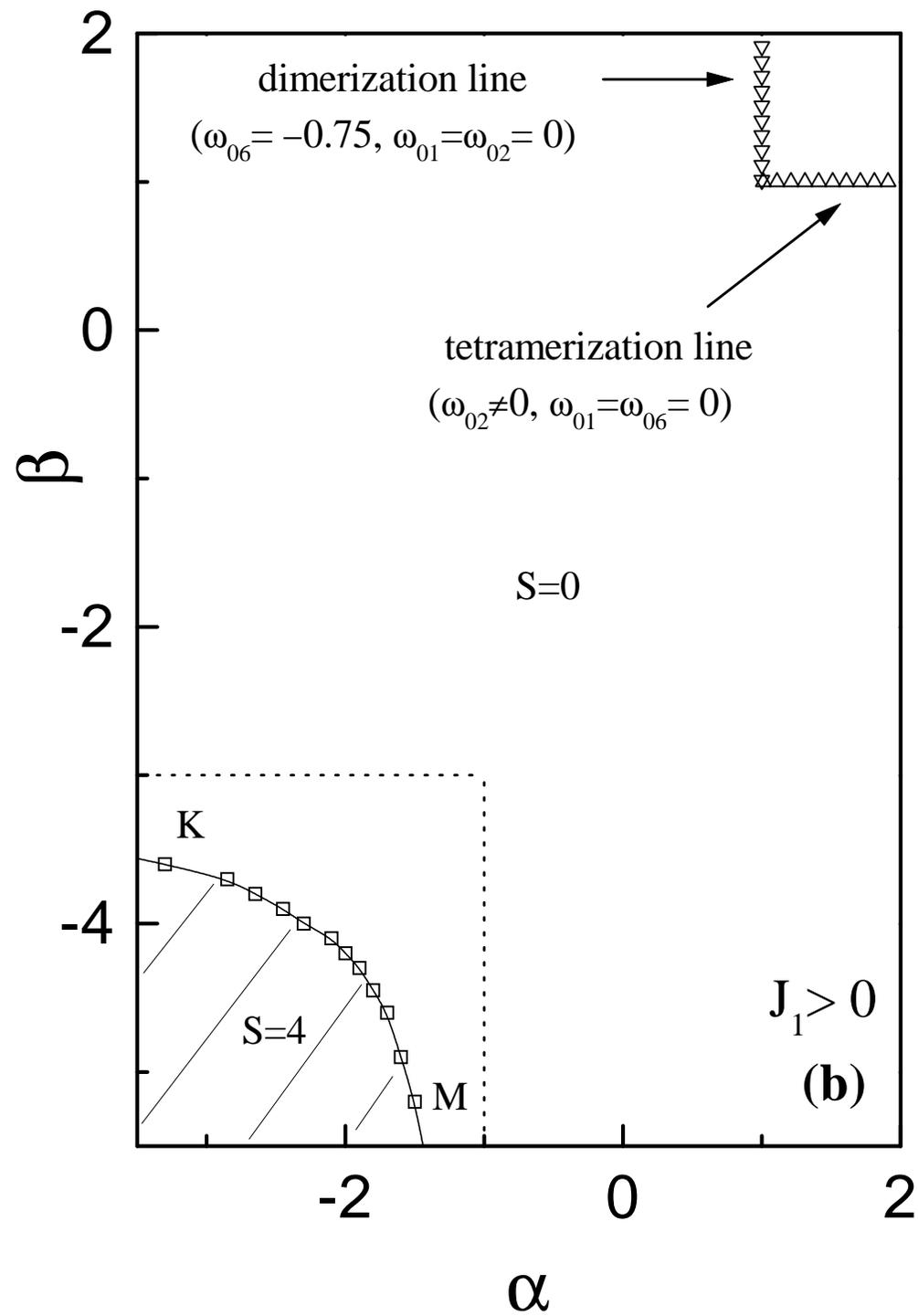
\end{document}